\documentclass[%
 aip,
 amsmath,amssymb,
 reprint,%
]{revtex4-1}

\usepackage{graphicx}
\usepackage{afterpage} 
\usepackage{dcolumn}
\usepackage{bm}

\usepackage[utf8]{inputenc}
\usepackage[T1]{fontenc}
\usepackage{mathptmx}
\usepackage{etoolbox}
\usepackage{xcolor}

\makeatletter
\def\@email#1#2{%
 \endgroup
 \patchcmd{\titleblock@produce}
  {\frontmatter@RRAPformat}
  {\frontmatter@RRAPformat{\produce@RRAP{*#1\href{mailto:#2}{#2}}}\frontmatter@RRAPformat}
  {}{}
}%
\makeatother

\begin{document}

\preprint{AIP/123-QED}

\title[Sample title]{ High performance vacuum  annealed $\mathrm{\beta}$-(Al\textsubscript{x}Ga\textsubscript{1-x})\textsubscript{2}O\textsubscript{3}/Ga\textsubscript{2}O\textsubscript{3} HFET with  f\textsubscript{T}/f\textsubscript{MAX} of 32/65 GHz}
\author{Chinmoy Nath Saha\textsuperscript{\dag}}

\author{Noor Jahan Nipu\textsuperscript{\dag}}



\author{Uttam Singisetti} 

 \affiliation{ Electrical Engineering, University at Buffalo, Buffalo, New York 14240, USA. \\
 \textsuperscript{\dag} These authors contributed equally to this work. }

 \email{uttamsin@buffalo.edu}

\date{\today}

\begin{abstract}
This letter reports high performance $\mathrm{\beta}$-(Al\textsubscript{x}Ga\textsubscript{1-x})\textsubscript{2}O\textsubscript{3}/Ga\textsubscript{2}O\textsubscript{3} Heterostructure FET (HFET) with improved regrowth process and successful Al\textsubscript{2}O\textsubscript{3} passivation. Highly scaled I shaped gate (100- 200 nm  L\textsubscript{G}) have been fabricated with degenerately doped (N++) source/drain contact regrown by ozone molecular beam epitaxy (MBE). Aiming to address the limitations observed in previous generation devices, this work incorporates a low-power BCl\textsubscript{3}/Ar and SF\textsubscript{6}/Ar plasma etching process to remove the AlGaO barrier layer and Ga\textsubscript{2}O\textsubscript{3} layer respectively before regrowth.
Additionally, the surface was cleaned and vacuum annealing was carried out before MBE regrowth to reduce any interface resistance between highly doped regrowth N++ and 2DEG layer. These meticulously designed fabrication steps enabled us to achieve the high DC current 0.5 A/mm at 5V drain bias with 6.1 $\mathrm{\Omega}$.mm on resistance (R\textsubscript{ON}) at V\textsubscript{GS}=3V, peak transconductance (g\textsubscript{m}) of 110 mS/mm at room temperature (V\textsubscript{DS}=15V) and around 0.8 A/mm (V\textsubscript{DS}=5V) peak I\textsubscript{ON} at low temperature (100K). Current gain cut-off frequency (f\textsubscript{T}) of 32 GHz and peak power gain cut-off frequency (f\textsubscript{MAX}) of 65 GHz were extracted from RF measurements. The reported f\textsubscript{MAX} is one of the highest for Ga\textsubscript{2}O\textsubscript{3} based RF devices to the best of authors’ knowledge. f\textsubscript{T}.L\textsubscript{G} product was estimated to be 6.1 GHz-$\mu$m for 191 nm L\textsubscript{G} and 32 GHz f\textsubscript{T}, which is one of the highest reported among Ga\textsubscript{2}O\textsubscript{3} devices. After passivating the device with thicker Al\textsubscript{2}O\textsubscript{3}, device shows no current collapse demonstrating first time successful traps passivation with Al\textsubscript{2}O\textsubscript{3} for $\mathrm{\beta}$-Ga\textsubscript{2}O\textsubscript{3} devices.

\end{abstract}
\maketitle

Ultrawide bandgap  $\mathrm{\beta} $-Ga\textsubscript{2}O\textsubscript{3} (Ga\textsubscript{2}O\textsubscript{3}) has attracted researchers attention worldwide due to its material  properties\cite{green2022beta} such as high critical electric field\cite{green20163}, good saturation velocity\cite{ghosh2017ab} resulting in high Johnson's figure of merit compared to GaN and SiC. Based on these encouraging prospects, $\mathrm{\beta} $-Ga\textsubscript{2}O\textsubscript{3} is a strong candidate for next-generation high-power RF amplifiers and switching applications. In the recent years, multi KV $\mathrm{\beta} $-Ga\textsubscript{2}O\textsubscript{3} diodes\cite{wang2023beta,mudiyanselage2021wide,farzana2021vertical} and FETs\cite{arkka1,kalarickal2021beta,dryden2022scaled} have been reported to show the feasibility for high power applications upto 10 KV. In 2017, $\mathrm{\beta}$-(Al\textsubscript{x}Ga\textsubscript{1-x})\textsubscript{2}O\textsubscript{3}/Ga\textsubscript{2}O\textsubscript{3} heterostructure was successfully demonstrated by krishnamoorthy et al.\cite{krishnamoorthy2017modulation} to utilize 2 dimensional electron gas (2DEG) at the hetero-structure and achieve better RF performance by enhancing mobility and reducing ionized impurity scattering. Simultaneous achievement of good RF performance and high breakdown voltage is necessary to implement Ga\textsubscript{2}O\textsubscript{3} in high power RF applications. In recent years, advancements in $\mathrm{\beta} $-Ga\textsubscript{2}O\textsubscript{3} technology have demonstrated significant progress in RF devices\cite{zhang2018,Zhanbo2019,zhou20231,Zhou2024,yu2023heterointegrated}. Additionally, our group has also achieved state-of-art RF performance by reporting 30 GHz 
 f\textsubscript{T} for AlGaO/GaO HFET\cite{vaidya2021enhancement} and  55 GHz f\textsubscript{MAX} for thin channel  $\mathrm{\beta} $-Ga\textsubscript{2}O\textsubscript{3} MOSFET\cite{saha2024thin} and 5.4 MV/cm average breakdown field\cite{sahabeta2}. We also analyzed DC-RF dispersion study in details for AlGaO/GaO HFET  and reported decent RF performance upto 250\textsuperscript{0}C\cite{saha2022temperature}. 

 One of the critical stages of $\mathrm{\beta}$-(Al\textsubscript{x}Ga\textsubscript{1-x})\textsubscript{2}O\textsubscript{3}/Ga\textsubscript{2}O\textsubscript{3} HFET fabrication is to etch the AlGaO layer and some unintentionally doped (UID) Ga\textsubscript{2}O\textsubscript{3} layer before doing the regrowth. In our previous reported HFET\cite{vaidya2021enhancement}, we found that high power ICP RIE based etch before the regrowth caused high interface resistance between N++ regrowth and the 2DEG, thereby limiting our transconductance (g\textsubscript{m}) and RF performance. Joishi et al \cite{joishi2020deep} demonstrated that MBE annealing at vacuum can recover some portion of damage caused by plasma etch. Arkka et al. \cite{bhattacharyya2021multi} reported low power SF\textsubscript{6}/Ar based plasma etching with etch rate 1-2 nm/min can cause very low damage to the interface and they found very low contact resistance. Combining both vacuum annealing and low power SF\textsubscript{6}/Ar can result in lower interface resistance between N++ regrowth layer and 2DEG, which can improve both DC and RF performance. The effect of thick Al\textsubscript{2}O\textsubscript{3} layer as a passivation dielectric to reduce/eliminate current collapse has not been explored for $\mathrm{\beta} $-Ga\textsubscript{2}O\textsubscript{3} devices. 

 In this letter, we have fabricated  $\mathrm{\beta}$-(Al\textsubscript{x}Ga\textsubscript{1-x})\textsubscript{2}O\textsubscript{3}/Ga\textsubscript{2}O\textsubscript{3} HFET using 100-200 nm L\textsubscript{G}. The device shows no non-linearity unlike our previous generation HFET, 0.5 A/mm on current and record transconductance upto 80-100 mS/mm which proves the effectivity of the developed low-power etching and MBE annealing at ultra-high vacuum. From RF measurement, we extracted f\textsubscript{MAX} of 65 GHz and  f\textsubscript{T} 32 GHz with high f\textsubscript{T}.L\textsubscript{G} product. Successful Al\textsubscript{2}O\textsubscript{3} passivation was demonstrated with no current collapse at higher V\textsubscript{DG,q} quiescent bias points.

\begin{figure}
\centerline{\includegraphics[width=1.0\columnwidth]{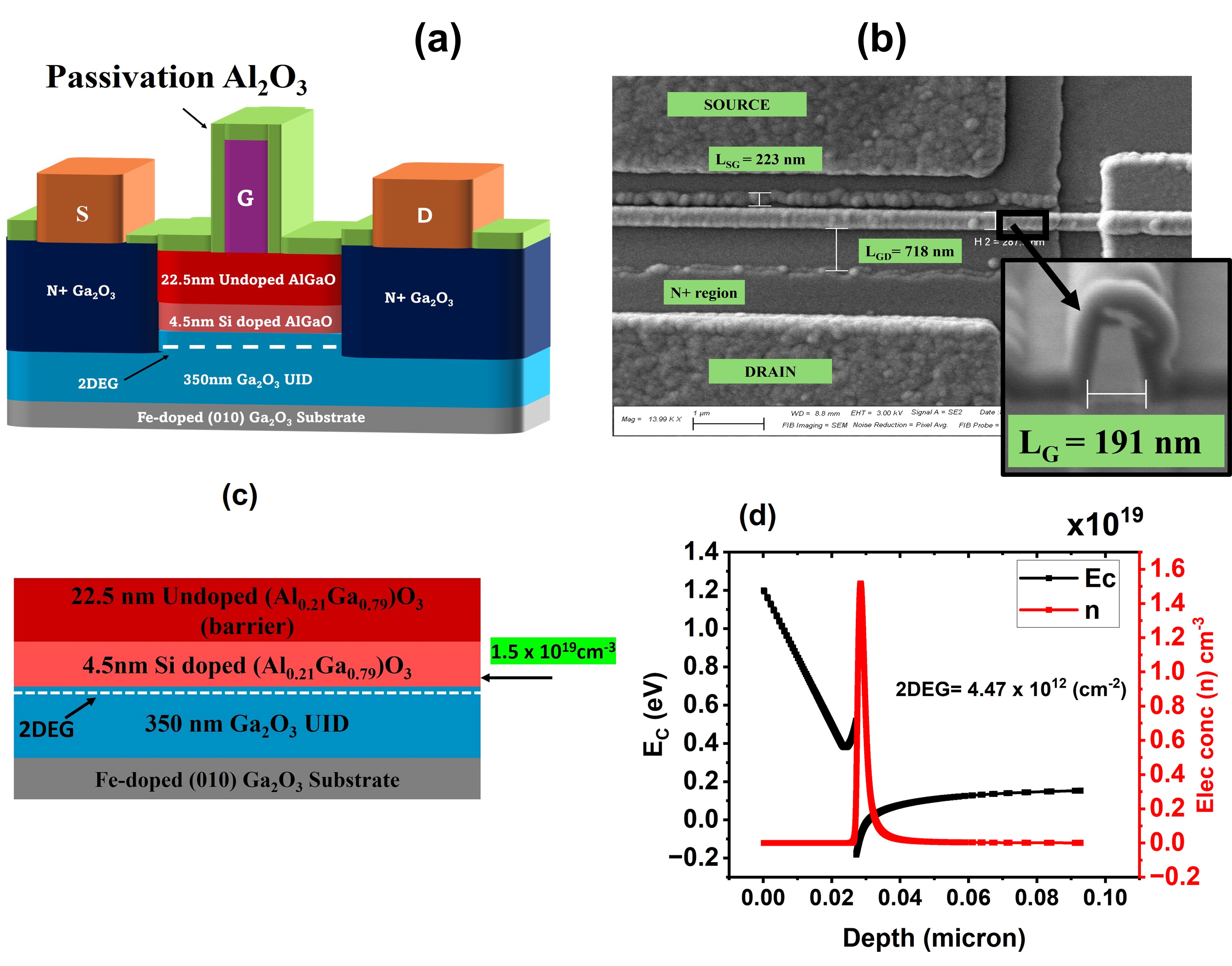}}
\vspace*{-2mm}
\caption{(a) Cross section of final fabricated device structure with Al\textsubscript{2}O\textsubscript{3} passivation (b) Magnified SEM and FIB image showing a completed I gate device with dimensions L\textsubscript{SG}/L\textsubscript{G}/L\textsubscript{GD}= 223 nm/191 nm/718 nm (c) Epitaxial stack of highly doped HFET. Here Si doping in AlGaO layer is 1.5 x 10\textsuperscript{19}cm\textsuperscript{-3}. (d) Simulated energy band diagram using the epitaxial stack. The extracted theoretical 2DEG charge density  is 4.47 x 10\textsuperscript{12}cm\textsuperscript{-2}.}
\label{fig1}
\end{figure}

The cross-section of the epitaxial structure of our device is shown in fig. \ref{fig1} (c) which is grown by Ozone  Molecular Beam Epitaxy (MBE) courtesy of Novel Crystal technology. On top of Fe- doped insulating substrate, UID (Unintentional doped) buffer layer of $\mathrm{\beta} $-Ga\textsubscript{2}O\textsubscript{3}
with 350 nm thickness was grown. Later, 4.5 nm Si doped (1.5 × 10\textsuperscript{19} cm\textsuperscript{-3}) $\mathrm{\beta}$ (Al\textsubscript{0.21}Ga\textsubscript{0.79})\textsubscript{2}O\textsubscript{3} layer and 22.5 nm $\mathrm{\beta}$-(Al\textsubscript{0.21}Ga\textsubscript{0.79})\textsubscript{2}O\textsubscript{3} undoped barrier layer were grown. The details growth method of HFET is reported by Vaidya et al. \cite{vaidya2019structural} in an earlier work. The simulated electron density from a self-consistent Schrodinger-Poisson solver using Silvaco
TCAD is shown in figure \ref{fig1}(d). The calculated 2DEG density is approximately 4.47  × 10\textsuperscript{12} cm\textsuperscript{-2}.

The initial steps of the device fabrication process are similar to our previous reported HFET in 2021.\cite{vaidya2021enhancement}.  In this work, we have used low power BCl\textsubscript{3}/Ar etch (ICP = 300, RIE= 50) to etch the AlGaO layer. Low power etch made sure that we removed AlGaO layer slowly and stopped on UID Ga\textsubscript{2}O\textsubscript{3} layer without doing much plasma damage. Later very low power  SF\textsubscript{6}/Ar etch (ICP =150, RIE= 50) was used with etch rate 1-2 nm/min. Immediately before MBE  regrowth, we have done a surface treatment with 1:3 HCl: DI water for 15 min to remove any interface contaminant. After that, sample was heated to 600\textsuperscript{0}C and annealed at ultra-high vacuum (10\textsuperscript{-10} torr) for 1 hour at the MBE growth chamber. Then 100 nm highly doped N++ layer was grown with a
doping density of 3 x 10\textsuperscript{19}cm\textsuperscript{-3} in the same MBE machine. Subsequently, regrowth mask removal and MESA isolation were carried out. We have deposited source-drain contact (Ti/Au/Ni = 50/120/30 nm) to form ohmic contact using e-beam evaporation. Later we deposited Ni/Au = 30 nm/180 nm for schottky gate formation. Finally, device was passivated using 100 nm Al\textsubscript{2}O\textsubscript{3} deposited by ALD at 300\textsuperscript{0}C. The final cross-section of the device schematic is shown in figure \ref{fig1}(a) with magnified SEM and FIB image (figure  \ref{fig1}(b)).

\begin{figure}
\centerline{\includegraphics[width=1.0\columnwidth]{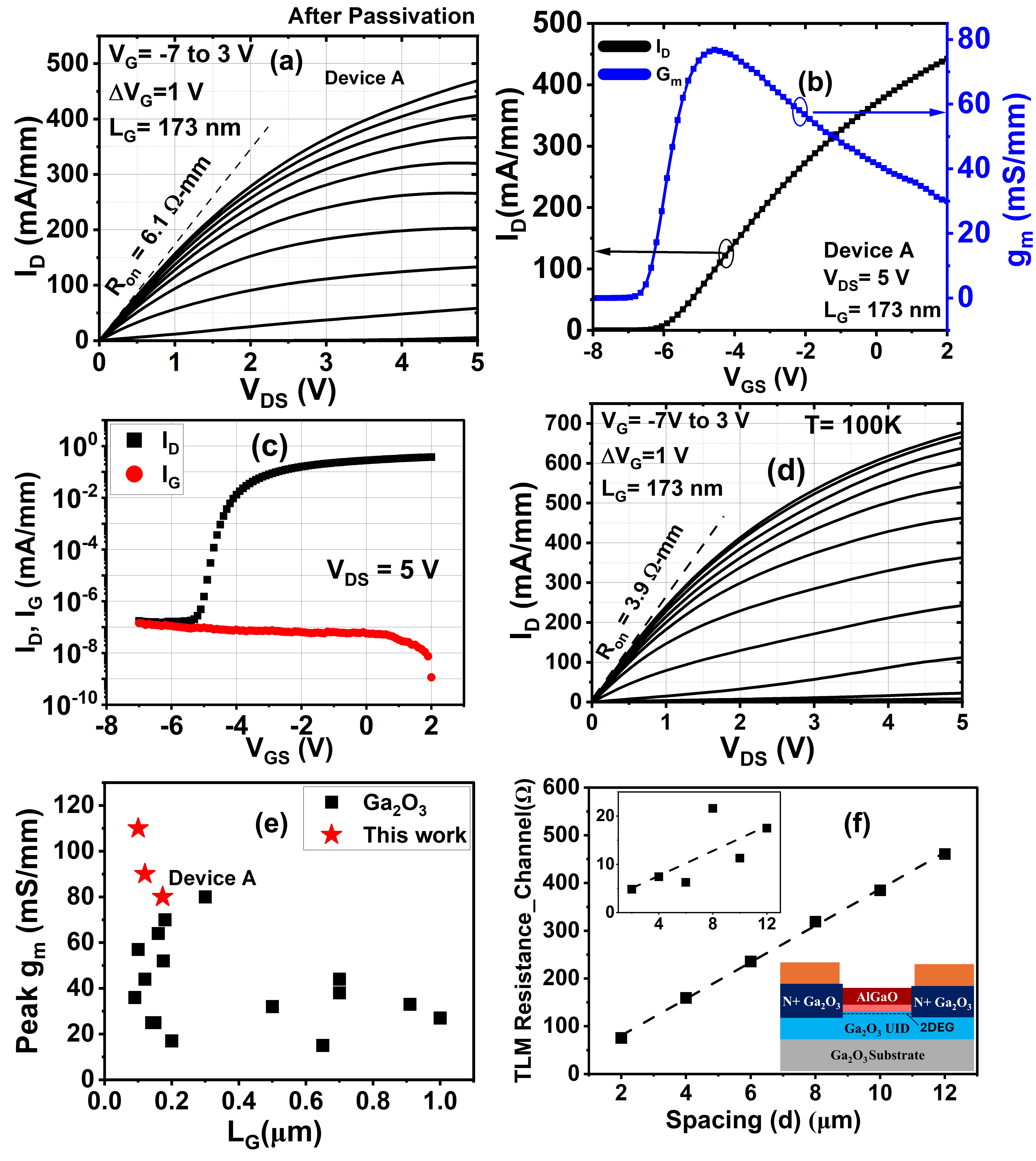}}
\vspace*{-2mm}
\caption{(a) I\textsubscript{D}-V\textsubscript{DS} output curve of device A after passivation at room temperature showing peak I\textsubscript{D}= 500 mA/mm (b) I\textsubscript{D}-V\textsubscript{GS} transfer curve of device A,(c) Semilog transfer curve showing I\textsubscript{D} and I\textsubscript{G}, (d) Output curve at T = 100 K showing peak I\textsubscript{D}= 0.8 A/mm with R\textsubscript{ON} = 3.9 $\Omega$.mm at V\textsubscript{GS}= 3V, (e) g\textsubscript{m} vs L\textsubscript{G} bench-marking of Ga\textsubscript{2}O\textsubscript{3} FETs from this work with other reported works, (f) TLM plot of channel resistance as a function of contact spacing, inset: TLM plot for the N++ doped region, showing resistance as a function of contact spacing and the cross-section image of TLM structure on N++ layer through the 2DEG layer.}
\label{fig2}
\end{figure}

We have measured the sheet resistance and contact resistance using the TLM (Transfer Length Method) structures which have been fabricated on the wafer. In figure \ref{fig2}(f), we have shown the TLM resistance with TLM spacing for both N++ regrowth layer and the access region connecting 2DEG. From the TLM structure on N++ layer through the 2DEG layer, we calculated total contact (R\textsubscript{C,total}) and sheet resistance of the access region connecting 2DEG (figure \ref{fig2}(f)). We extracted total contact resistance (R\textsubscript{C,total}) of 0.5 $\mathrm{\Omega}$ mm  with sheet resistance (R\textsubscript{sheet, ch}) around 9.6 K$\mathrm{\Omega}$/$\mathrm{\square}$. Using this R\textsubscript{sheet} and predicted charge density of 4.47 x 10\textsuperscript{12} cm\textsuperscript{2}, the mobility is estimated to be 145 cm\textsuperscript{2}/vs, which is higher than our previous gen. HFET\cite{vaidya2021enhancement}. However, the sheet resistance is a lot lower than our previously reported HFET due to the higher doping used in this HFET sample. Again, as calculated from figure \ref{fig2}(f) inset plot, the TLM structure on N++ regrowth layer gave around 0.35 $\mathrm{\Omega}$ mm lateral contact resistance (R\textsubscript{C,N++}) and a sheet resistance (R\textsubscript{sheet,N++}) of 316 $\mathrm{\Omega}$/$\mathrm{\square}$. The contact resistance of N++ MBE grown layer is higher than the MOCVD regrown  N++ layer reported in our MOSFET \cite{sahabeta2} due to lower doping in the MBE N++ layer (3 x 10\textsuperscript{19} cm\textsuperscript{-3}) compared to  MOCVD regrowth doping (1 x 10\textsuperscript{20}cm\textsuperscript{-3}) \cite{alema2022low}.

The HFETs were characterized at room temperature before and after passivation. The DC characteristics were measured using 4155B semiconductor parameter analyzer. From the output characteristics we extracted maximum drain current (I\textsubscript{D,max}) = ~500 mA/mm and on resistance (R\textsubscript{ON}) = 6.1 $\Omega$ mm at  V\textsubscript{GS} = 3 V, and V\textsubscript{DS} = 5 V from the DUT of 173 nm gate length (L\textsubscript{G}) (figure \ref{fig2}(a)). The I\textsubscript{D}-V\textsubscript{GS} transfer curve of the device is showing peak g\textsubscript{m} of 80 mS/mm at V\textsubscript{DS}= 5V for L\textsubscript{G} = 173 nm with threshold voltage -6V (figure \ref{fig2}(b)). We extracted threshold voltage of -4V to -6V by doing a device map across the wafer. The negative threshold voltage is coming from the low work function of Ni gate. We also plotted the semilog transfer curve (figure \ref{fig2}(c)) which shows I\textsubscript{D} and gate current I\textsubscript{G}. From the I\textsubscript{D} plot we extracted > 10\textsuperscript{6} on-off ratio. The gate current is around 10\textsuperscript{-8} mA/mm  throughout the V\textsubscript{GS} range which is lower than our previous reported HFET\cite{vaidya2021enhancement}. A benchmark plot of peak g\textsubscript{m} with gate length in figure \ref{fig2}(e) is showing a comparison of our peak g\textsubscript{m} of this work with other $\mathrm{\beta} $-Ga\textsubscript{2}O\textsubscript{3} FETs, where we reported peak g\textsubscript{m} of 110 mS/mm of device B of L\textsubscript{G} 100 nm at V\textsubscript{DS} = 15 V (supplementary materials). We can see that top two peak g\textsubscript{m} have been reported from this work.



Also, from the output characteristics (figure \ref{fig2}(a)) we did not observe any non-linearity at lower V\textsubscript{DS} as reported in previous HFET \cite{vaidya2021enhancement}. In our 1st gen. HFET, non-linearity probably resulted from high-power ICP-RIE etch at the regrowth interface. Since we do not see this behavior in this sample, we can fairly conclude that our process optimization including low power etch, HCl dip before regrowth and MBE annealing helped to remove the non-linearity and improved I\textsubscript{D}. We observed no significant shift in threshold voltage and no collapse for peak I\textsubscript{D} and peak g\textsubscript{m} after passivation from measuring I\textsubscript{D}- V\textsubscript{GS} transfer curves of all devices of the wafer proving the effectiveness of our thicker Al\textsubscript{2}O\textsubscript{3} scheme. Si\textsubscript{3}N\textsubscript{4} passivation in our earlier HFET\cite{saha2022temperature} caused a shift in threshold voltage possibly attributed to high-temperature deposition or PECVD plasma damage. We also observed slight current collapse even after Si\textsubscript{3}N\textsubscript{4} passivation in our thin channel MOSFET \cite{sahabeta2}. In comparison, Al\textsubscript{2}O\textsubscript{3} passivation in this work seems to be successful based on DC analysis.

Additionally, R\textsubscript{ON} reduced almost 5 times compared to our last work. Higher 2DEG density and a successful process optimization enabled this high current density with low R\textsubscript{ON}. Also, no RTA was performed for this sample which can also explain why contact resistance is higher for MBE sample compared to MOCVD regrowth samples. 

A low-temperature DC measurement is also carried out using lakeshore cryo probe station along with 4155B analyzer stepping from the 10 K to 350 K with 50 K steps. We achieved high peak I\textsubscript{D} of around 800 mA/mm with very low R\textsubscript{ON} = 3.9 $\Omega$-mm at V\textsubscript{GS}= 3 V and V\textsubscript{DS}= 5 V at 100 K (figure \ref{fig2}(d)). This on current is about 60\% higher than room temperature peak I\textsubscript{D}. An increase of channel mobility due to the reduced effect of optical phonon scattering at low temperature is causing this higher current \cite{zhang2019evaluation}. At 100 K we also achieved high peak g\textsubscript{m} of 135 mS/mm at V\textsubscript{DS}= 5 V (supplementary materials) with no shift in threshold voltage from room temperature value. 

\begin{figure}
\centerline{\includegraphics[width=1.0\columnwidth]{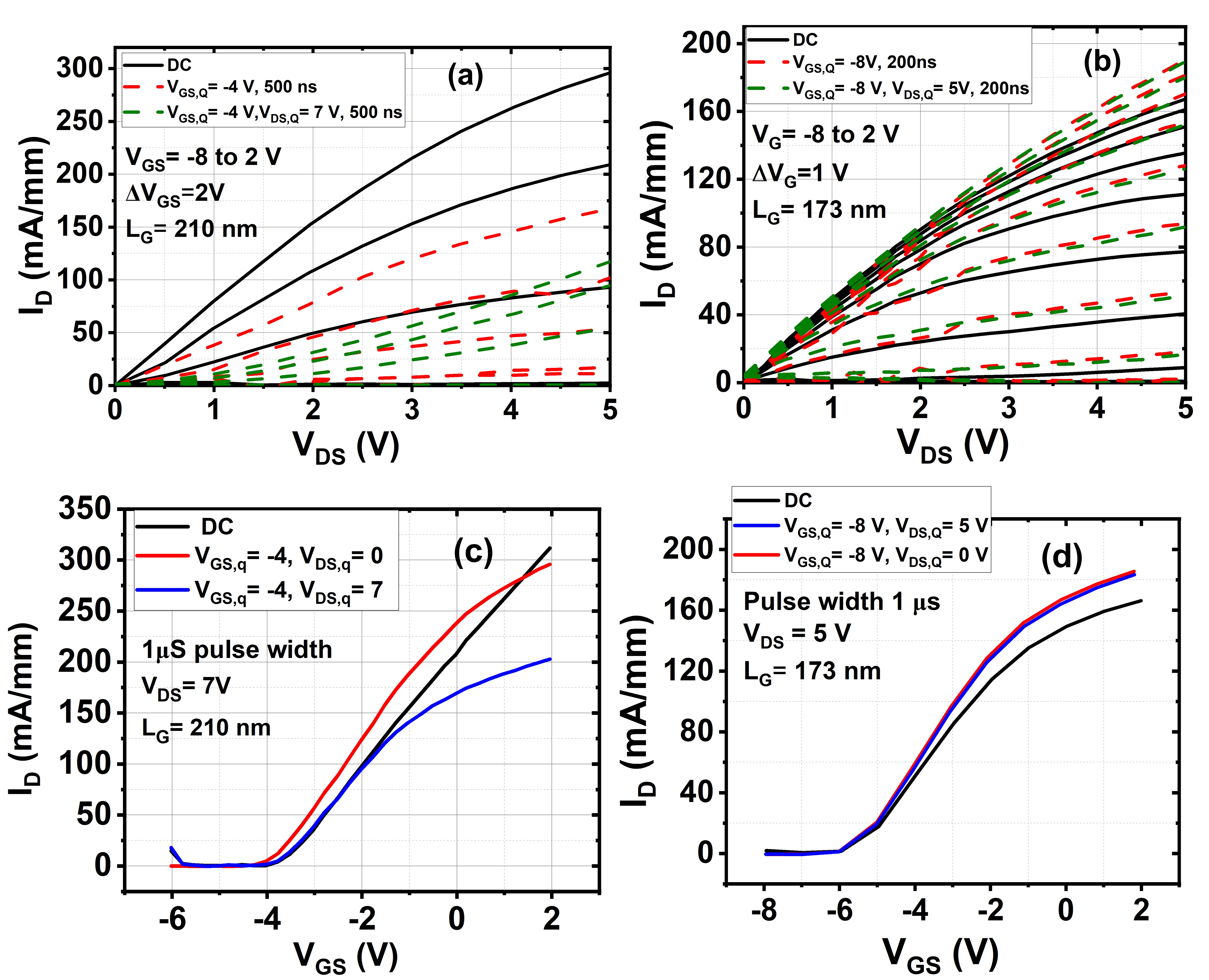}}
\vspace*{-2mm}
\caption{(a) I\textsubscript{D}-V\textsubscript{DS} output characteristics before passivation, showing severe current collapse under both gate pulse and double-pulse conditions, (b) Output characteristics after passivation, demonstrating no current collapse under gate and double pulse condition at 200 ns pulse width, (c) I\textsubscript{D}-V\textsubscript{GS} transfer characteristics before passivation, showing current collapse and shift in V\textsubscript{TH}, (d) Transfer characteristics after passivation, showing no current collapse and no shift in V\textsubscript{TH} under all double-pulse conditions.}
\label{fig3}
\end{figure}

Initially, we carried out pulsed IV measurements using auriga system before passivation. We measured I\textsubscript{D}- V\textsubscript{DS} output curve for different quiescent bias conditions and compared with DC measurements (figure \ref{fig3}(a)). Drain pulse condition (V\textsubscript{DS,q} = 0) has a higher current compared to DC (supplementary materials) which rules out any traps in the buffer region\cite{saha2022temperature,vaidya2021temperature}. But for gate pulse condition (V\textsubscript{GS,q} = -4 V), 140 mA/mm current drop was observed compared to DC. Based on our previous works\cite{vaidya2021temperature,saha2022temperature,sahabeta2} and GaN literature\cite{meneghesso2013trapping,meneghesso2004surface}, traps under the gate are the primary reason behind the current collapse of gate lag conditions. The current collapse is more severe (50\%) for the double pulse condition with V\textsubscript{GS,q} = -4 V and V\textsubscript{DS,q} = 7 V. This particular pulsed IV condition favors the trap in the gate-drain access region resulting in current collapse in our unpassivated device.

After passivation with 100 nm Al\textsubscript{2}O\textsubscript{3}, we measured the pulsed IV at three pulse conditions. 
For the output characteristics, after passivation and RF measurements, at 200 ns pulse width we found higher peak I\textsubscript{D} than DC for drain pulse (V\textsubscript{DS,q}= 0) (supplementary materials) and gate pulse (V\textsubscript{GS,q}= -8 V) condition. This particular device has V\textsubscript{TH}= -6 V. So we used V\textsubscript{GS,q}= -8 V as shown in figure \ref{fig3}(b) for the gate pulse condition. The same scenario is also evident for the dual pulse condition (V\textsubscript{GS,q} = -8 V, V\textsubscript{DS,q}= 5 V) (figure \ref{fig3}(b)) where we found higher peak I\textsubscript{D} than DC. Other devices in the sample also show no DC-RF dispersion after passivation. We can conclude from this analysis that, Al\textsubscript{2}O\textsubscript{3} has passivated the traps and eliminated DC-RF dispersion in AlGaO/GaO HFET. This is the first demonstration of successful passivation by Al\textsubscript{2}O\textsubscript{3} for $\mathrm{\beta}$-Ga\textsubscript{2}O\textsubscript{3}.  Recently Dryden et al. \cite{dryden2022scaled} reported similar passivation technique for $\mathrm{\beta}$-Ga\textsubscript{2}O\textsubscript{3} MESFET, but they found moderate dispersion after passivation.

I\textsubscript{D} - V\textsubscript{GS} transfer curve using different quiescent bias conditions is an effective way to understand the location of traps. Figure \ref{fig3}(c) shows I\textsubscript{D} - V\textsubscript{GS} transfer curve before passivation, where at V\textsubscript{GS,q} = -4 V and V\textsubscript{DS,q} = 0 V bias condition negligible current collapse is found, but we observe a small shift in threshold voltage. Theoretically, this bias condition should favor trapping of the electron under gate \cite{meneghesso2013trapping}. Combined use of a high drain voltage (V\textsubscript{DS,q} = 7 V) and negative gate voltage (V\textsubscript{GS,q} = -4 V) resulted in a significant decrease (33\%) of peak drain current at V\textsubscript{GS} = 2 V. This observation implies the existence of traps in the gate-drain access region which caused current collapse.
 
After passivation, we observed no shift in threshold voltage and no current collapse for both negative gate bias (V\textsubscript{GS,q} = -8 V, V\textsubscript{DS,q} = 0 V)  and combined use of high drain voltage and negative gate voltage (V\textsubscript{GS,q} = -8 V, V\textsubscript{DS,q} = 5 V) quiescent bias points as evident from figure \ref{fig3}(d). Proper surface passivation after Al\textsubscript{2}O\textsubscript{3} deposition can be the preliminary reason for the absence of current collapse since traps are passivated. We noticed no current collapse for other devices of the sample at higher drain bias (V\textsubscript{DS,q} = 5 V) even for 200 ns pulse width (supplementary materials).
 
\begin{figure}[t]
   \centerline{\includegraphics[width=1.0\columnwidth]{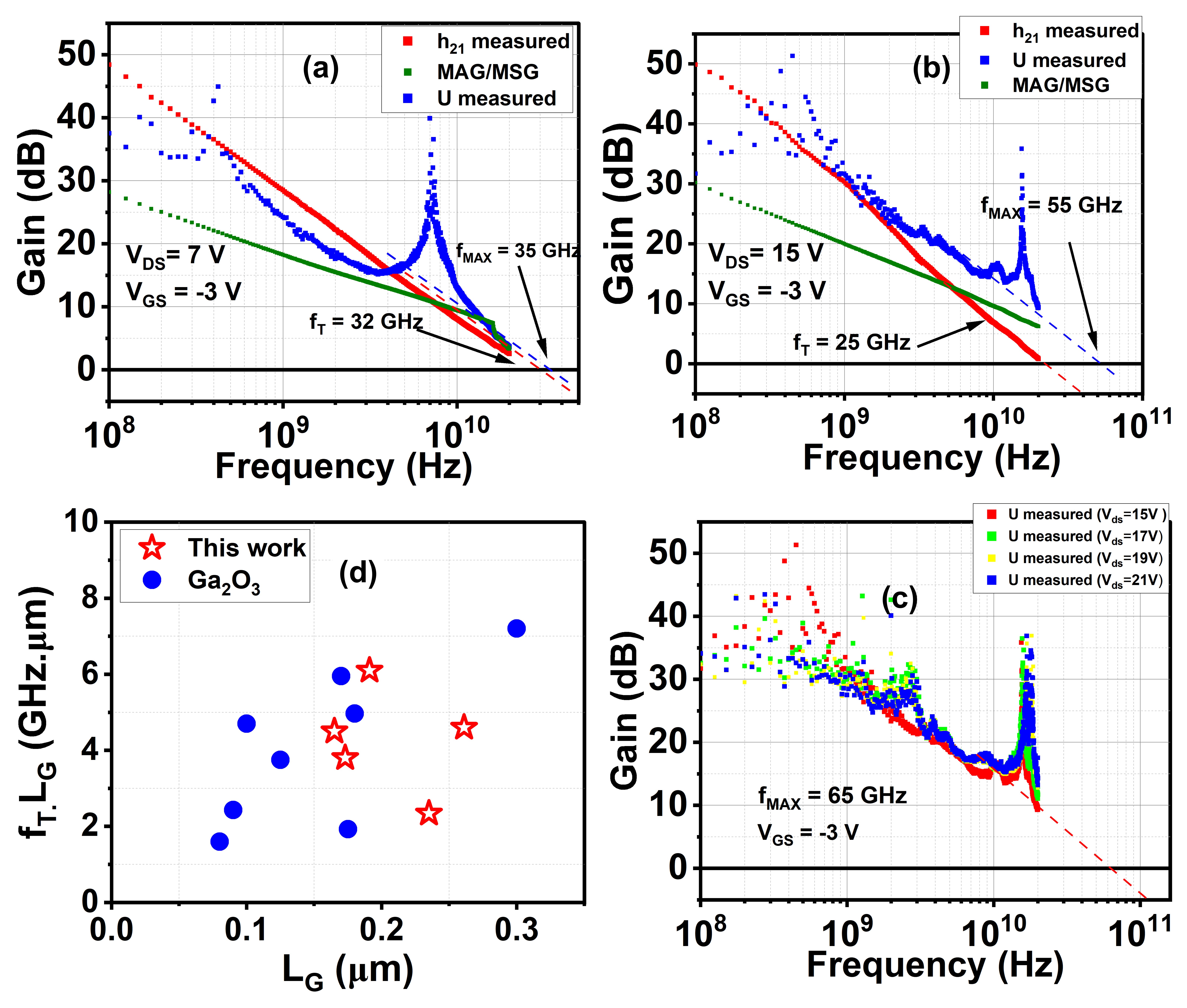}}
   \caption{(a) Small signal RF measurement showing f\textsubscript{T} = 32 GHz at V\textsubscript{GS} = -3 V, V\textsubscript{DS} = 7 V (b)  f\textsubscript{MAX} = 55 GHz at V\textsubscript{GS} = -3 V, V\textsubscript{DS} = 15 V for L\textsubscript{G} = 191 nm,(c) Variation of  f\textsubscript{MAX} with V\textsubscript{DS} showing 65 GHz f\textsubscript{MAX} at V\textsubscript{DS}=21 V for L\textsubscript{G} = 191 nm, (d)  f\textsubscript{T}. L\textsubscript{G} product with different L\textsubscript{G} benchmark plot comparison of our measured f\textsubscript{T}. L\textsubscript{G} product across the wafer comparing with other reported Ga\textsubscript{2}O\textsubscript{3} RF FETs }
    \label{fig4}
\end{figure}

RF performance was analyzed using ENA 5071C vector network analyzer from 100 MHz to 20 GHz. We have measured short circuit current gain (h\textsubscript{21}), unilateral current gain (U), and MAG/MSG  and extrapolated them to 0 dB. An isolated open-pad structure on the device wafer was measured to de-embed the parasitic pad capacitance \cite{koleeandembed}. Surprisingly before passivation, we did not achieve any gain. Huge DC-RF dispersion or current collapse before passivation at ns pulse width can be attributed to this phenomenon. After passivation,  we extracted f\textsubscript{T} = 32 GHz at V\textsubscript{GS} = -3 V, V\textsubscript{DS} = 7 V (figure \ref{fig4}(a)) and  f\textsubscript{MAX} = 55 GHz at V\textsubscript{GS} = -3 V, V\textsubscript{DS} = 15 V (figure \ref{fig4}(b)) from gate length 0.191 $\mu$m. 

\begin{figure}[htbp]
    \centerline{\includegraphics[width=1.0\columnwidth]{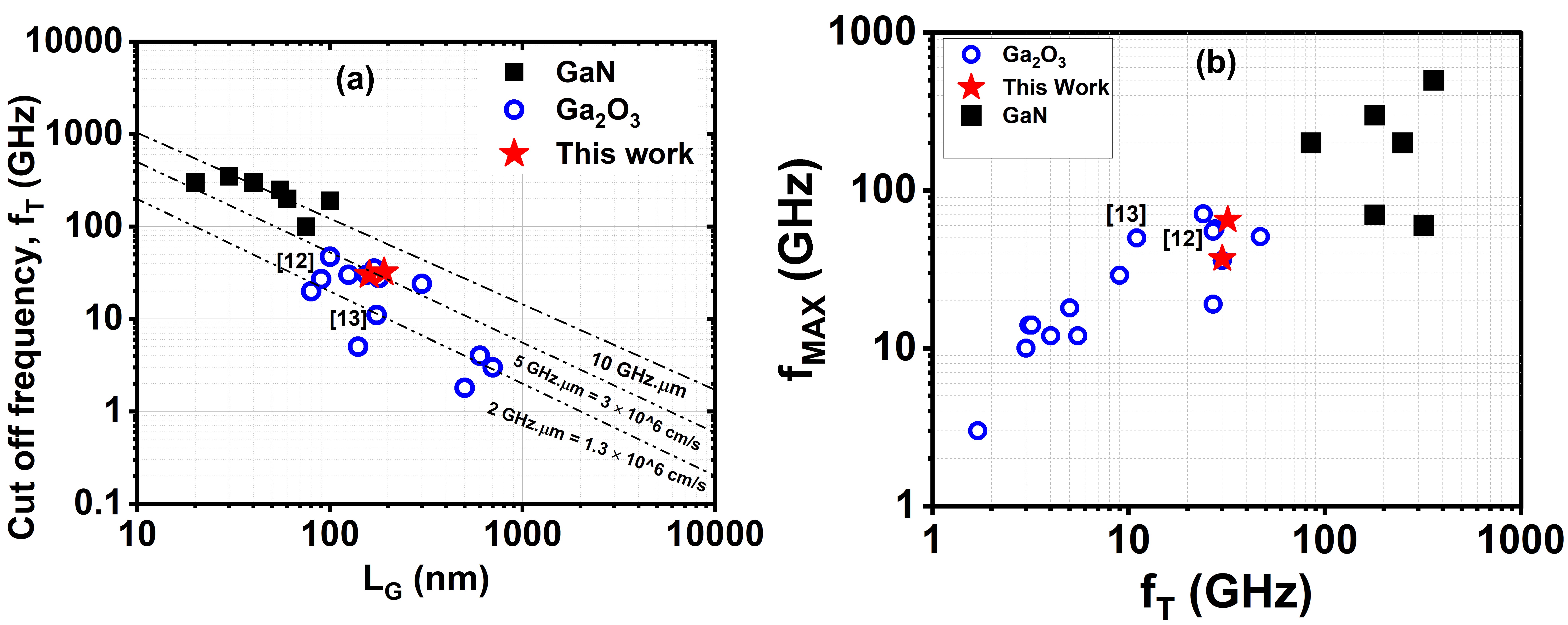}}
   \caption{(a) f\textsubscript{T} vs L\textsubscript{G} benchmark plot of our devices compared to other $\mathrm{\beta}$-Ga\textsubscript{2}O\textsubscript{3}  devices in literature and AlGaN/GaN HEMTs (b) f\textsubscript{T} vs f\textsubscript{MAX} scatter plot showing our devices have one of the highest  f\textsubscript{MAX} reported in literature among $\mathrm{\beta}$-Ga\textsubscript{2}O\textsubscript{3} FETs.}
    \label{fig5}
\end{figure}

It corroborates our claim that Al\textsubscript{2}O\textsubscript{3} passivates the traps and DC-RF dispersion is eliminated. This is the first demonstration of an improvement in RF performance after passivation for $\mathrm{\beta}$-Ga\textsubscript{2}O\textsubscript{3}  using Al\textsubscript{2}O\textsubscript{3} as a passivation layer. Figure \ref{fig4}(c) shows the increment of  f\textsubscript{MAX} with increasing V\textsubscript{DS} due to gate-drain feedback capacitance \cite{Green2017,chabak2018}. We have achieved around 65 GHz f\textsubscript{MAX} at V\textsubscript{DS} = 21 V along with a demonstration of better stability by possessing higher gain with increasing drain bias. Figure \ref{fig4}(d) shows the measured f\textsubscript{T}.L\textsubscript{G} products with different L\textsubscript{G} measured across the wafer. Here we have choosen the highest f\textsubscript{T}.L\textsubscript{G} product among other values which is 6.1 GHz.$\mu$m for L\textsubscript{G}= 0.191 $\mu$m and 32 GHz f\textsubscript{T}. It also shows the benchmark plot comparison of our FET with other $\mathrm{\beta}$-Ga\textsubscript{2}O\textsubscript{3} RF FETs which shows that our calculated f\textsubscript{T}.L\textsubscript{G} product is one of the highest values reported for $\mathrm{\beta}$-Ga\textsubscript{2}O\textsubscript{3} FETs. This value corresponds to  V\textsubscript{sat} approximately 3 x 10\textsuperscript{6} cm/s (figure \ref{fig5}(a)). We also plotted f\textsubscript{T} vs f\textsubscript{MAX} scatter plot comparing both f\textsubscript{T} and f\textsubscript{MAX} of our devices to the literature showing our f\textsubscript{T} and f\textsubscript{mAX} are one of the highest among other $\mathrm{\beta}$-Ga\textsubscript{2}O\textsubscript{3} FETs (figure \ref{fig5}(b)). Besides, f\textsubscript{T} and f\textsubscript{MAX} as a function of V\textsubscript{GS} follow transfer curve trend (supplementary materials). 


In summary, we have achieved state-of-the-art DC and RF performance for our AlGaO/GaO HFET using optimized surface cleaning, low-power etch, and MBE annealing. Our device reports I\textsubscript{ON}$~$500 mA/mm for DC, and on/off ratio > 10\textsuperscript{6} simultaneously. We achieved record  peak g\textsubscript{m} of 110 mS/mm, f\textsubscript{MAX} 65 GHz with 6.1 GHz.$\mu$m f\textsubscript{T}.L\textsubscript{G} product. No current collapse has been observed after successful Al\textsubscript{2}O\textsubscript{3} passivation with high RF performance. Our device demonstrates that beta gallium oxide has encouraging future prospects for next-generation RF applications.

\section*{SUPPLEMENTARY MATERIAL}
See the supplementary material for interface resistance measurement, I\textsubscript{D}-V\textsubscript{D} output and  I\textsubscript{D}-V\textsubscript{G} transfer curve before passivation, theoritical calculation of f\textsubscript{T}, Breakdown measurement, pulsed I\textsubscript{D}-V\textsubscript{D} output curve for drain pulse, I\textsubscript{D}-V\textsubscript{G} transfer curve and  transconductance (before passivation, after passivation and after stress) for 200 ns Pulse width, and f\textsubscript{T} and f\textsubscript{MAX} as a function of V\textsubscript{GS}.

\begin{acknowledgments}
We acknowledge the support from AFOSR (Air Force Office of Scientific Research) under award FA9550-18-1-0479 (Program Manager: Ali Sayir), from NSF under awards  ECCS 2019749, 2231026 from Semiconductor Research Corporation under GRC Task ID 3007.001, and Coherent II-VI Foundation Block Gift Program. This work used the electron beam lithography system acquired through NSF MRI award ECCS 1919798.

\end{acknowledgments}

\section*{Data Availability Statement}

The data that support the findings of this study are available from the corresponding author upon reasonable request.


\section{REFERENCES}
\nocite{*}
\bibliography{aipsamp}

\end{document}